\documentclass[prc,aps,amsfonts,amsmath,superscriptaddress,floatfix,twocolumn,linenumbers,nofootinbib]{revtex4}
\usepackage{graphicx,float} 
\usepackage{epsfig} 
\usepackage{rotating}
\usepackage{mwe}    % loads �blindtext� and �graphicx�
\usepackage{braket}
\usepackage{physics}    
\usepackage[colorlinks=true, linkcolor=blue, citecolor=blue, urlcolor=blue]{hyperref}
\usepackage{graphicx}
\usepackage{subcaption}
\usepackage{ulem}
\usepackage{tikz}

\def\ga{\,\,\raise0.14em\hbox{$>$}\kern-0.76em\lower0.28em\hbox
{$\sim$}\,\,}
\def\la{\,\,\raise0.14em\hbox{$<$}\kern-0.76em\lower0.28em\hbox
{$\sim$}\,\,}

\begin{document}

\title{Modeling Ultra-High-Energy Cosmic Rays propagation using the input from Configuration Interaction Shell Model}

\author{O. Le Noan}
\email{oscar.lenoan@iphc.cnrs.fr}
\affiliation{Universit\'e de Strasbourg, IPHC, 23 rue du Loess 67037 Strasbourg, France\\
CNRS, UMR7178, 67037 Strasbourg, France}
\author{S. Goriely}
\affiliation{Institut d'Astronomie et d'Astrophysique, Universit\'e Libre de Bruxelles, Campus de la Plaine, CP-226, 1050 Brussels, Belgium}
\author{E. Khan}
\affiliation{IJCLab, Universit\'e Paris-Saclay IN2P3-CNRS, F-91406 Orsay Cedex, France}
\affiliation{Institut Universitaire de France (IUF)}
\author{K. Sieja}
\email{kamila.sieja@iphc.cnrs.fr}
\affiliation{Universit\'e de Strasbourg, IPHC, 23 rue du Loess 67037 Strasbourg, France\\
CNRS, UMR7178, 67037 Strasbourg, France}

\begin{abstract} 
\begin{description}
\item[Background] The dipole response of a nuclear system, characterized by its photon strength function (PSF), is a key ingredient of many applications of nuclear structure, ranging from nuclear reactor design and nuclear waste transmutation to astrophysical models of nucleosynthesis and stellar evolution. While the majority of those applications require the knowledge of PSF of mid-mass and heavy nuclei, there  is now renewed interest in $E1$ strength distributions of light nuclei in the framework of the PANDORA project, which aims at an understanding of the mass distribution of ultrahigh-energy cosmic radiation (UHECR).
UHECR is of extragalactic origin and its interaction along the travel path is dominated by photoabsorption of cosmic background radiation boosted to the Giant Dipole Resonance (GDR) energy region in the center-of-mass system.
Thus, systematic knowledge of the photoabsorption cross sections in light nuclei and of their subsequent particle decay is required. 
\item[Purpose] The purpose of this work is to enhance the database of available theoretical evaluations of PSF of light nuclei that are necessary in the studies of UHECR propagation.   
\item[Methods] We employ the Configuration Interaction Shell Model (CI-SM) approach to provide predictions of $E1$ dipole response for $p$ and $sd$-shell nuclei, with mass number $A$ between 7
and 40. Theoretical predictions are compared to available data and to existing predictions from phenomenological and microscopic models. Finally, the impact
of using of CI-SM PSF on the predicted propagation of a $^{40}$Ca UHECR source is studied.
\item[Results] As the CI-SM approach captures all the nucleonic correlations in a given configuration space, the CI-SM predictions of the centroids and widths of PSF exhibit stronger variation and fragmentation, respectively, than those from linear-response and from phenomenological models. When used in the nuclear reaction network they lead to an outcome that is compatible with
the recently developed RQFAMz approach while are at stronger variance with the predictions employing the D1M+QRPA approach.
\item[Conclusions] The CI-SM method is shown to be a valuable alternative to provide PSF for the modeling of the propagation of the UHECR, so far only achieved with linear-response models and phenomenological models of nuclear strength functions. While this work covers a considerable fraction of nuclei necessary in such applications (namely $p$ and $sd$-shell nuclei), future work will also treat $pf$-shell systems and hence complete the CI-SM systematics of dipole PSF necessary for such applications.

\end{description}

\end{abstract}
 
%\pacs{21.60.-n}

\date{\today}

\maketitle
\section{Introduction}
\label{intro}
The average response of a nucleus to an electromagnetic perturbation is characterized by its photon strength function (PSF). PSFs are key quantities in many areas of nuclear physics, serving as essential tools for probing nuclear structure, from single-particle motion to collective excitations, as well as in the description of astrophysical processes. Among all types and multipolarities of electromagnetic response, the electric dipole (E1) response is usually the most probable, especially in the $\gamma$-ray region of 10-30 MeV encompassing the well-known giant dipole resonance which is a fully collective dipole mode with protons and neutrons oscillating out of phase \cite{harakeh_giant_2001}. In neutron-rich nuclei, the oscillation of the excess neutrons against the isospin-saturated core is predicted to generate a low-lying E1 strength, referred to as the pygmy dipole resonance (PDR) \cite{zilges_pygmy_2023}. Studying the PDR provides a means to probe the neutron-skin thickness in medium to heavy nuclei \cite{tamii_complete_2011, piekarewicz_pygmy_2011}, places constraints on the nuclear symmetry energy \cite{roca-maza_neutron_2015}, and provides insight into the properties of neutron stars \cite{horowitz_neutron_2001}. PSFs are also key ingredients to the calculations of radiative neutron capture cross sections, which are needed e.g. to model the intermediate \cite{Martinet24,Wiedeking25} or rapid \cite{Arnould07,Arnould2020,Cowan2021} neutron capture processes of nucleosynthesis. Last but not least, PSFs are necessary in the simulations of the propagation of ultra-high energy cosmic rays (UHECRs). This topic has recently gained significant attention through the PANDORA project (Photo-Absorption of Nuclei and Decay Observation for Reactions in Astrophysics) \cite{tamii_pandora_2022}, which was initiated to study the photo-nuclear reactions of light nuclei both experimentally and theoretically. Part of this initiative focuses on improving the characterization of UHECR's photo-interactions, a crucial component in the theoretical modeling of their propagation through extragalactic space. UHECR photo-interactions can significantly vary depending on the chosen PSF inputs in the calculations of photo-nuclear cross sections \cite{kha05, Kido2021}. To investigate the propagation of UHECR, PSFs of nuclei with mass $A \le 60$ are needed \cite{tamii_pandora_2022}. 
 Experimental photodata are scarce and often mutually contradictory in this target mass region. Therefore, one must resort to theoretical predictions of PSFs. Large scale PSF predictions primarily rely on parameterized models like the simple modified Lorentzian model (SMLO) \cite{plujko_giant_2018, goriely_simple_2019} and on declinations of linear response approaches such as the quasi-particle random phase approximation (QRPA) \cite{Paar2007, goriely_gogny-hfbqrpa_2018} and its quasi-particle finite amplitude method (QFAM) formulation \cite{Nakatsukasa2007, Nakatsukasa2011}, which has the advantage of being computationally cheaper than the former. Provided that phenomenological adjustments, such as mass dependent centroid shift and folding width, are added, such microscopic predictions are efficient for mid-mass and heavy nuclei, which display mainly smooth bulk-like responses. On the other hand, in the case of light nuclei, linear response fails to reproduce the PSF fragmentation \cite{le_noan_electric_2025}. To have reliable PSFs predictions for light nuclei, one may rely on microscopic approaches that account for correlations beyond the harmonic approximation of the QRPA, such as second QRPA \cite{gambacurta_low-lying_2011}, QRPA plus phonon coupling \cite{litvinova_mode_2010}, projected generator coordinate method \cite{kimura_structure_2017, Bofos_2025}, or the Configuration Interaction Shell Model (CI-SM)\cite{caurier_shell_2005}. The large-scale evaluations of PSF within those models are, however, not yet accessible due to numerical complexity. In particular, calculations of electric dipole excitation strength functions within the CI-SM framework are relatively scarce in the literature, mainly due to the challenges associated with deriving multi-shell valence-space interactions and the large dimensions of the Hamiltonian matrices involved. Recently, the $E1$ PSFs of all long-lived $sd$-shell nuclei have been calculated within the CI-SM framework \cite{le_noan_electric_2025}. Building on these efforts, the present study aims at: 1) expanding the CI-SM investigation through a systematic analysis of the $E1$ photoabsorption strength function in $p$ and $sd$-shell nuclei; 2) benchmarking the results against experimental data; 3) comparing to existing theoretical predictions, and 4) estimating the impact of the CI-SM  predictions on the propagation of UHECRs.\\
The present work is organized as follows: we describe the details of the CI-SM calculations performed in this work in Sec. \ref{THEO}. In Sec. \ref{COMP} a comparison of our theoretical results to experimental data and other available models of PSF is presented, both for nuclei close to the stability line and 
in the neutron-rich region for O and Ne chains. The impact of model differences on modeling of the photodisintegration of the UHECR are discussed in Sec. \ref{UHECR}. 
Finally, the conclusions of this study are collected in Sec. \ref{CONC}.

\section{Shell Model approach to $E1$ strength calculations\label{THEO}}

\subsection{Theoretical framework}
The CI-SM \cite{caurier_shell_2005}, or large-scale shell-model approach, allows one to diagonalize the (typically) one- and two-body nuclear Hamiltonian in a configuration space generated by arranging nucleons within a chosen set of single-particle orbits, known as the model space. 
The shell-model Hamiltonian reads:
\begin{equation}
H=\sum_i \epsilon_i c_i^\dagger c_i+ \sum_{ijkl}V_{ijkl}c_i^\dagger c_j^\dagger c_l c_k+\beta H_{\textrm{c.m.}}
\end{equation}
where the center-of-mass (c.m.) Hamiltonian, scaled by a coefficient $\beta = 10$, is included to shift the spurious c.m. eigenvalues into an energy range outside the scope of the present study \cite{gloeckner_spurious_1974}.
The $E1$ strength is generated by the isovector dipole operator:
\begin{equation}
\hat O_{1\mu}=-e\frac{Z}{A}\sum_{i=1}^N r_i Y_{1\mu}(\hat r_i)+e\frac{N}{A}\sum_{i=1}^Z r_iY_{1\mu}(\hat r_i).
\label{oper}
\end{equation}
Shell-model calculations performed in this work span masses between $A= 7$ and $A= 40$, 
which means $p$-shell and $sd$-shell nuclei. This requires two distinct model spaces to achieve the description of $E1$ excitations of these systems. 
Our approach to derive the $E1$ strength of $sd$-shell nuclei, using the PSDPF interaction \cite{bouhelal_psdpf_2011}, was recently presented in detail in Ref. \cite{le_noan_electric_2025}. To compute the $E1$ response of $p$-shell nuclei we used the WBP effective Hamiltonian defined in the in the $s-p-sd-pf$ model space \cite{warburton_effective_1992}. In both model spaces, we allow only for $0\hbar \omega$ ground states (GS) and $1\hbar \omega$ excited states. The reduced transition probability in CI-SM is calculated as 
\begin{equation}
B_{\nu 0}=\frac{1}{2J_0+1}\langle \nu||\hat O|| 0\rangle^2,
\end{equation}
where $\ket{0}$ is the nuclear GS, $\ket{\nu}$ is an excited eigenstate, $J_0$ is the GS total angular momentum quantum number and $\hat O$ is the transition operator, in our case from Eq. \ref{oper}. The $B(E1)$ distributions are obtained through the Lanczos strength-function method, enabling an efficient determination of the strength per energy interval \cite{caurier_shell_2005}.
The selection of the starting vector, known as the pivot, in the Lanczos diagonalization procedure is arbitrary. 
Given a transition operator $\hat O$ one can define a pivot of the form $\hat O |0\rangle$, known as the sum rule state and carry on Lanczos diagonalization.
The unitary matrix $U$ that diagonalizes the Hamiltonian after $N$ Lanczos iterations contains, then, in its first row, the necessary amplitudes $\langle \nu|\hat O| 0\rangle$. Such calculations were carried out using the $m$-scheme shell-model code ANTOINE \cite{caurier_shell_2005} with 300 iterations, which ensures the convergence of the PSFs. 

The $k^{\textrm{th}}$ moment of a $B(E1)$ distribution is given by the sum rule of order $k$:
\begin{equation}
S_k=\sum_\nu E_{\textrm{exc}}(\nu)^k B_{\nu 0},
\label{SRk}
\end{equation}
where $E_{\textrm{exc}}(\nu)$ is the excitation energy of the $\ket{\nu}$ eigenstate. In terms of these moments, the centroid and width are defined as: 
\begin{eqnarray}
\bar S=\frac{S_1}{S_0},\quad \Delta S=\sqrt{\frac{S_2}{S_0}-\bar S^2}.
\label{centrowidtheq}
\end{eqnarray}
To phenomenologically account for the coupling to the continuum, we additionally fold $B(E1)$ distributions with a generalized Lorentzian: 
\begin{equation}
S(E)=\sum_\nu B_{\nu 0}\frac{2}{\pi}\frac{\Gamma E^2}{\big (E^2-E_\textrm{exc}(\nu)^2 \big)^2+\Gamma^2 E^2},
\label{folding}
\end{equation}
which has been shown to provide the correct behavior at low excitation energy \cite{goriely_large-scale_2002, goriely_reference_2019, goriely_gogny-hfbqrpa_2018}. The width parameter is set to $\Gamma = 1$ MeV for $sd$-shell nuclei, with the exception of $^{40}$Ca, for which $\Gamma = 3.5 $ MeV is adopted to account for the bunching of the strength into a few $B_{0\nu}$ peaks, a consequence of the limited 
correlations due to the complete filling of the $sd$-shell. For $p$-shell nuclei, a width of $\Gamma = 3.5$ MeV is employed in all cases. We present the results of the computations as PSFs (in ~MeV$^{-3}$) defined by the relation \cite{bartholomew_gamma-ray_1973}: 
\begin{equation}
f_{E1}(E)=\frac{16\pi}{27(\hbar c)^3}S(E)\,.
\end{equation}

\subsection{Normalization of the PSF strength\label{norm}}
Systematic overestimation of the $E1$ total strength function appears to be a common feature with available shell-model calculations \cite{le_noan_electric_2025, orce_global_2023, sagawa_low_2001, sagawa_pigmy_1999, utsuno_photonuclear_2015, togashi_e1_2018}. As in the case of $M1$, $E2$, $E3$, and Gamow-Teller transitions \cite{caurier_shell_2005, bouhelal_shell_2017, matsubara_quenching_2021, ogunbeku_universal_2025, suhonen_value_2017}, the $E1$ operator must be renormalized, or “dressed,” to incorporate correlations beyond the model space. Microscopic methods for mapping an operator into a model space include Lee–Suzuki approach \cite{coraggio_perturbative_2020} and similarity renormalization group techniques \cite{parzuchowski_ab_2017}. Phenomenological effective charges are often favored over these methods due to their complexity and the difficulties associated with managing multi-shell model spaces. In Ref. \cite{le_noan_electric_2025}, we determined the $E1$ effective charges for $sd$-shell nuclei as $(e_{\textrm{eff}}^n, e_{\textrm{eff}}^p) = (-0.8Z/A, 0.8N/A)$, corresponding to an overall reduction of the total strength by a factor of $Q^2=0.64$. In this work, we adopt a slightly different value of $Q^2$ determined as:
\begin{equation}
    Q^2S_1 = S_{\textrm{\scriptsize TRK}}(1+\kappa),
\end{equation}
where $S_1$ is the energy-weighted sum rule (EWSR) from Eq. \ref{SRk} obtained using the bare dipole operator, $S_{\textrm{\scriptsize TRK}} = 14.8 NZ/A \quad e^2\textrm{fm}^2~\textrm{MeV}$ is the Thomas-Reich-Kuhn (TRK) sum rule, and $\kappa$ an enhancement factor \cite{mottelson_nuclear_1969}.  It is known  that the enhancement factor critically depends on an energy cutoff $E_\gamma^{\textrm{max}}$ \cite{harakeh_giant_2001, traini_study_1987, lipparini_sum_1989, ishkhanov_giant_2021}. Typically for $E_\gamma^{\textrm{max}} \sim 140$ MeV, $\mathcal{O}  (\kappa) = 1$ while if $E_\gamma^{\textrm{max}} \sim E_{\textrm{\scriptsize GDR}}$, $\kappa$ is much smaller. In the present calculations, we set the energy range of interest to $E_\gamma^{\textrm{max}} = 50$ MeV. A phenomenological adjustment comes in the enhancement factor assumed in the present calculations, as $\kappa = 0.14$ for $sd$-shell nuclei \cite{le_noan_electric_2025} and $\kappa = 0$ for $p$-shell nuclei. 

\section{Comparison to experimental data and to other models\label{COMP}}
\subsection{$p$-shell nuclei}
The electric dipole response of selected $p$-shell nuclei has been investigated previously within the framework of CI-SM with the WBP family of Hamiltonians in Refs. \cite{suzuki_electric_2004, orce_global_2023}. A systematic comparison of the dipole polarizability (proportional to the $k=-1$ moment of the $B(E1)$ distribution) with experimental data was performed in Ref.~\cite{orce_global_2023}. The well-known tendency of CI-SM calculations to overestimate the $E1$ response was observed, but remained unaccounted for. Contrary to those previous works, here we have performed systematic $E1$ response calculations for all long-lived ($\tau > 1$ hour) $p$-shell nuclei applying a phenomenological normalization of the operator, as explained in Sec. \ref{norm}.

\begin{figure*}[htbp]
    \begin{center}
 \includegraphics[width=\textwidth]{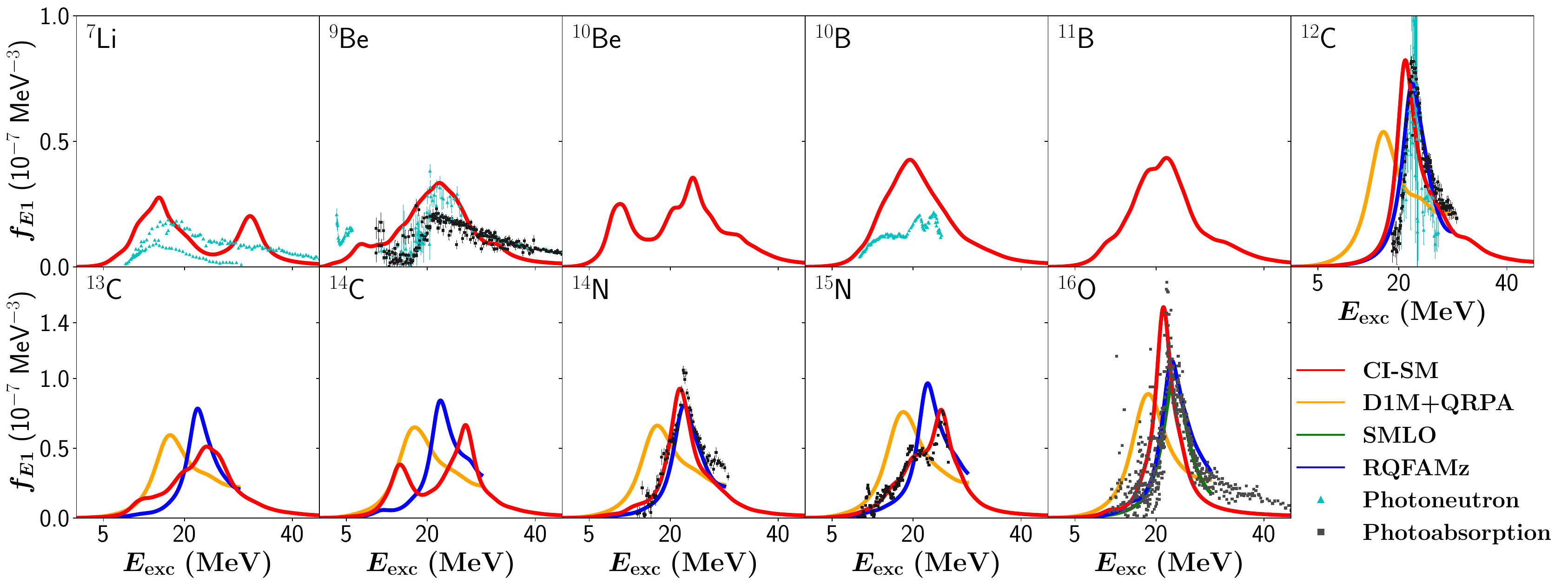}
\caption{Photoabsorption strength functions in $p$-shell nuclei. Results obtained in this work (red lines) are compared to available experimental data \cite{goriely_reference_2019} (symbols), RQFAMz results from \cite{gonzalez-miret_zaragoza_large-scale_2025} (blue lines), SMLO PSFs \cite{plujko_giant_2018, goriely_simple_2019} (yellow lines) and D1M+QRPA calculations \cite{goriely_gogny-hfbqrpa_2018}.}
\label{p}  
\end{center}
\end{figure*}

In Fig.~\ref{p}, we compare our CI-SM predictions to available photoabsorption data from Ref.~\cite{goriely_reference_2019}, to the state-of-the-art relativistic QFAM calculations from \cite{gonzalez-miret_zaragoza_large-scale_2025}, dubbed hereafter RQFAMz, to the Gogny-QRPA results of Ref.~\cite{goriely_gogny-hfbqrpa_2018}, dubbed hereafter D1M+QRPA, and to the phenomenological SMLO prediction \cite{plujko_giant_2018, goriely_simple_2019}, which is essentially a Lorentzian fit to available data. Both D1M+QRPA and RQFAMz predictions incorporate global phenomenological energy shifts of the dipole strength, allowing for a more accurate reproduction of both the centroids and the positions of the energy maxima. The D1M+QRPA peak is, however, shifted systematically towards lower energies with respect to the one from RQFAMz.  As seen in Fig.~\ref{p}, the overall agreement between our CI-SM results 
and photodata is satisfactory. It is worth noticing that in $^{15}$N the CI-SM predictions closely reproduce the fragmentation of the distribution, whereas the RQFAMz calculations yield a single-peak structure. This behavior is expected and illustrates how the CI-SM framework accounts for correlations beyond the linear-response–based approaches.

\subsection{$sd$-shell nuclei}
 
\begin{figure*}[htbp]
    \begin{center}
       \centering
       \includegraphics[width=\textwidth]{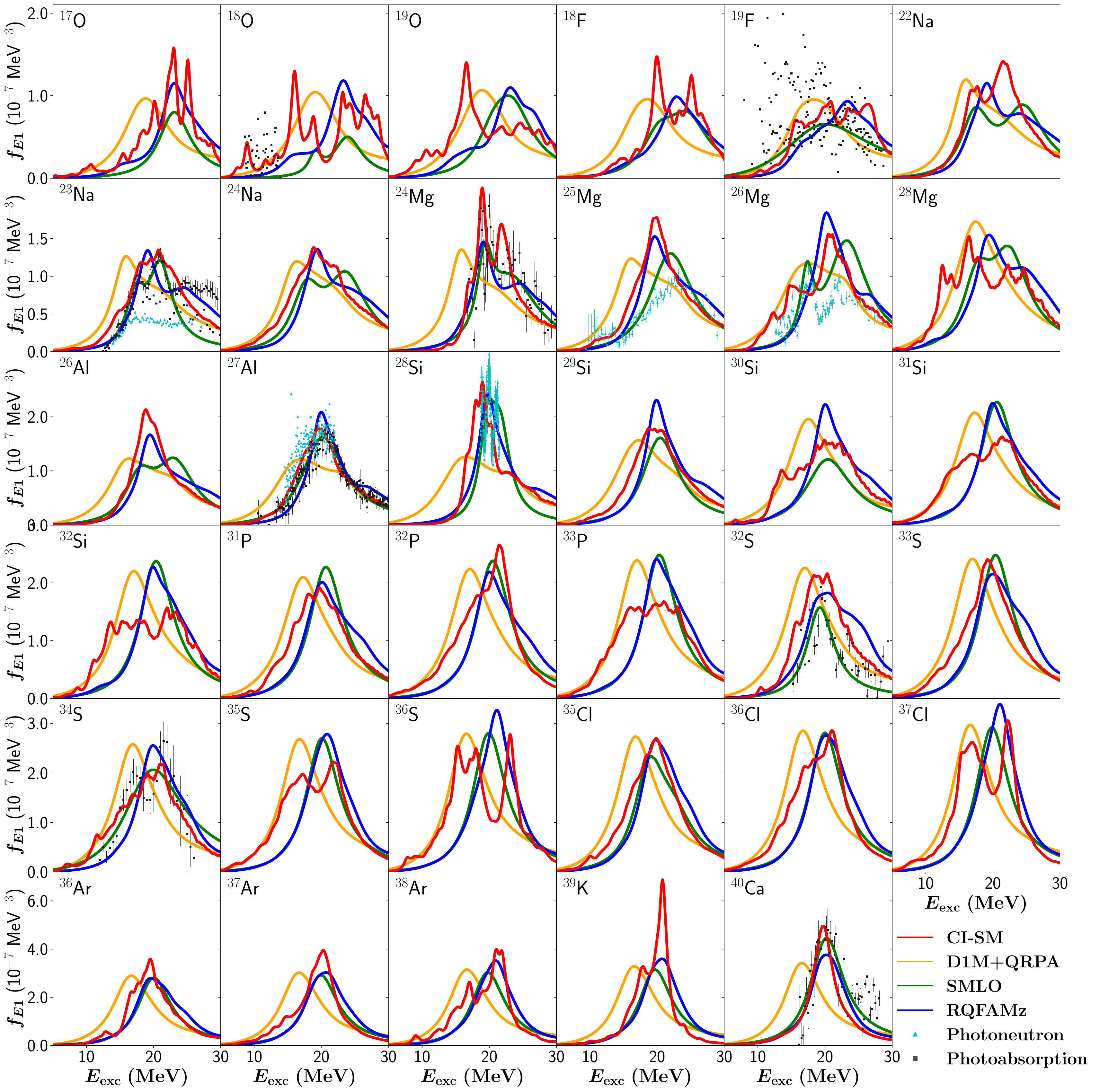}
\caption{Same as Fig.~\ref{p} for $sd$-shell nuclei}
%Photoabsorption strength functions of long-lived $sd$-shell nuclei obtained in this work (red lines) versus available experimental data \cite{goriely_reference_2019} (symbols), RQFAMz results from \cite{gonzalez-miret_zaragoza_large-scale_2025} (blue lines), SMLO PSFs \cite{plujko_giant_2018, goriely_simple_2019} (green lines) and D1M+QRPA calculations from \cite{goriely_gogny-hfbqrpa_2018} (yellow lines).}
\label{sd}
\end{center}
\end{figure*}

\begin{table}
\caption{Deviation from experiment for the three microscopic models, quantified by the mean error $\varepsilon$ and the root-mean-square deviation $\sigma$, in MeV.\label{tab-e} }
\begin{tabular}{ccccc} 
\hline 
\hline 
 {Models}  & {$\varepsilon_{\bar{S}}$} & {$\sigma_{\bar{S}}$} & {$\varepsilon_{\Delta S}$} & {$\sigma_{\Delta S}$} \\
\hline
     {CI-SM} & {$-0.36$} & {$0.85$} & {$0.27$} & {$0.51$} \\
     {D1M + QRPA} & {$-1.47$} & {$1.71$} & {$0.13$} & {$0.49$} \\
     {RQFAMz} & {$0.54$} & {$0.89$} & {$-0.33$} & {$0.48$}  \\

\hline
\hline 
\end{tabular}
\end{table}

\begin{figure}[htbp]
    \begin{center}
       \centering
    \begin{subfigure}[b]{0.24\textwidth}
        \includegraphics[width=\textwidth]{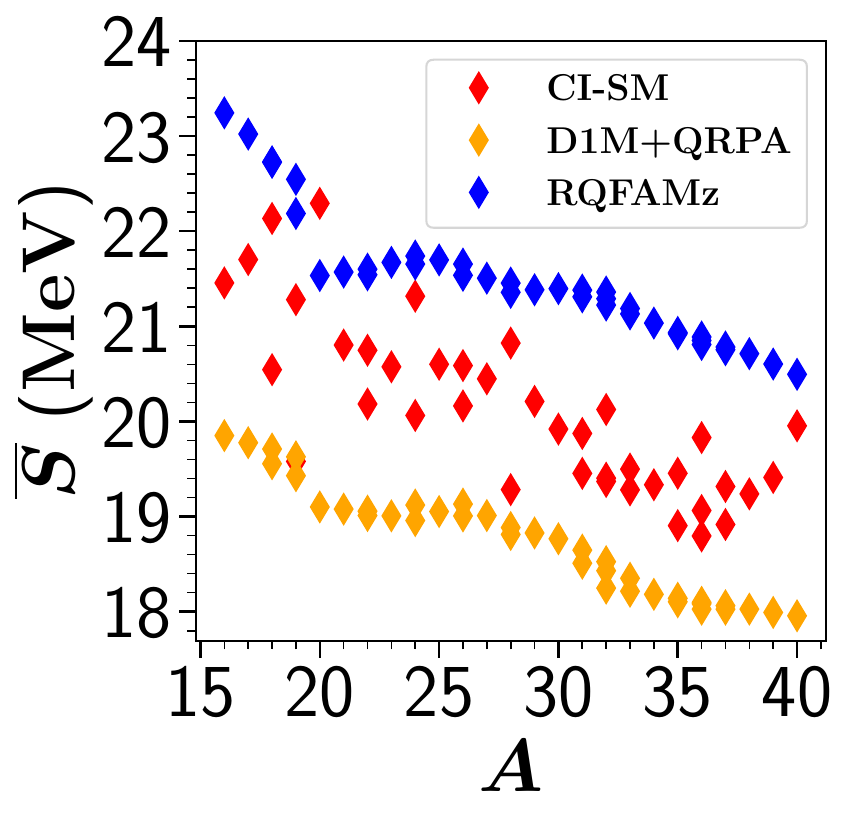}
            \begin{tikzpicture}[overlay]
        \node[anchor=north west] at (-0.65,4.2) {(a)};
    \end{tikzpicture}
    \end{subfigure}
    \hfill
        \begin{subfigure}[b]{0.23\textwidth}
        \includegraphics[width=\textwidth]{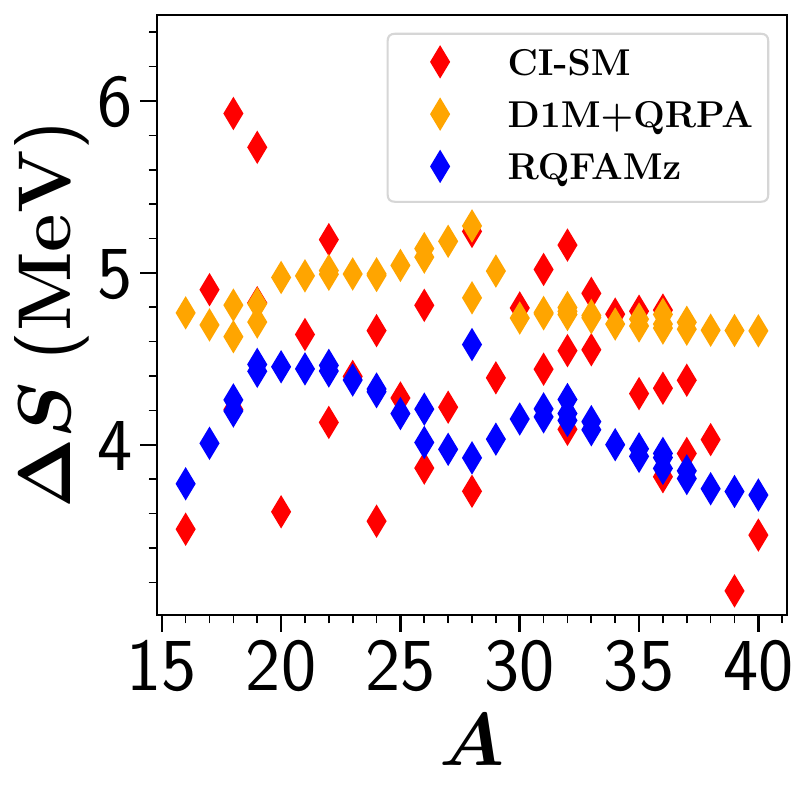}
                    \begin{tikzpicture}[overlay]
        \node[anchor=north west] at (-0.65,4.2) {(b)};
    \end{tikzpicture}
    \end{subfigure}

    \caption{Panel (a): centroids (up to $30$ MeV) of $E1$ distributions as a function of the mass number $A$ for long-lived $sd$-shell nuclei. Panel (b): width (up to $30$ MeV) of $E1$ PSFs as a function of the mass number $A$ for long-lived $sd$-shell nuclei. In red CI-SM, in orange D1M+QRPA \cite{goriely_reference_2019}, in blue RQFAMz results \cite{gonzalez-miret_zaragoza_large-scale_2025}. }    
\label{SDSsd}
\end{center}
\end{figure}
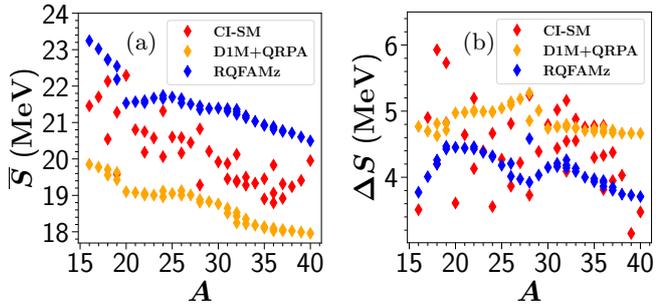

In Ref.~\cite{le_noan_electric_2025}, we presented systematic calculations of PSFs in long-lived $sd$-shell nuclei and compared them to available data and D1M+QRPA predictions from Ref.~\cite{goriely_gogny-hfbqrpa_2018}. In this work we extended the CI-SM calculations to provide predictions for all 141 existing isotopes that can be described in this valence space. Figs. \ref{sd} and \ref{SDSsd} compare our CI-SM results for $sd$-shell nuclei close to the stability line with those obtained with other models, namely D1M+QRPA, SMLO \cite{plujko_giant_2018, goriely_simple_2019} and RQFAMz \cite{gonzalez-miret_zaragoza_large-scale_2025}. It can be observed that, while in certain cases the CI-SM predictions are in reasonable agreement with the SMLO and RQFAMz PSFs, in many others the CI-SM results exhibit a higher degree of fragmentation and slight differences in strength magnitude, width, and centroid, which may influence photodisintegration cross sections. Fig. \ref{SDSsd} shows the evolution of the centroid and the width as a function of $A$. The decrease of the centroid energy with increasing mass is consistently observed across all three microscopic models with a similar trend. It is worth noting that both D1M+QRPA and RQFAMz include global phenomenological adjustments affecting not only the centroid position but also the folding width. The CI-SM centroids exhibit less smooth variations compared to linear-response-based approaches, as they can incorporate rapid changes in the shell structure and in the GS correlations. Similarly, the CI-SM widths are more scattered and show no clear systematic trend, whereas the D1M+QRPA and RQFAMz widths exhibit a smooth mass dependence. In Ref.~\cite{le_noan_configuration_2025}, we presented a systematic comparison of CI-SM centroids and widths with experimental data. The CI-SM centroid predictions appeared slightly lower than the measured values, with a root-mean-square (rms) deviation of $0.85$ MeV across the 26 cases (see Table \ref{tab-e}). For the CI-SM widths, the corresponding rms deviation is $0.50$ MeV, with no systematic under- or over-estimate\footnote{Note that the centroid and the width as defined in Eq. \ref{centrowidtheq} are independent of the normalization of the PSF strength described in Sec. \ref{norm}.}. In CI-SM, we account for the Landau damping as well as most of the spreading width; therefore, the calculated width is almost not sensitive to the folding, which only accounts for the escape width. For comparison, the rms deviation in D1M+QRPA predictions amounts to $1.71$ MeV on the centroids and $0.49$ MeV on the widths, while RQFAMz leads to an rms deviations of $0.89$ MeV on the centroids and $0.48$ MeV  on the widths. Hence, even without empirical adjustments to the centroids or widths, the CI-SM predictions remain slightly more accurate than RQFAMz PSFs for $sd$-shell nuclei. One may note that in many cases the experimental PSFs themselves are mutually inconsistent (see Fig. \ref{sd}), so these rms values should be interpreted with caution. We shall also recall that the shell-model PSDPF effective interaction \cite{bouhelal_psdpf_2011} was only adjusted to reproduce the low-energy spectroscopic data. Therefore, a possible improvement in the PSF predictions could be to incorporate in the fit of the interaction also the measured centroids and widths. From a purely practical perspective one could apply a simple centroid shift, in analogy with the procedure used in QFAM/QRPA calculations. Because the uncorrected CI-SM reproduces the known centroids better than QRPA-based models with empirical corrections, we chose to base our predictions for nuclei used in Sec. \ref{sec:uhecr} on the bare CI-SM results.

\subsection{O and Ne isotopic chains}
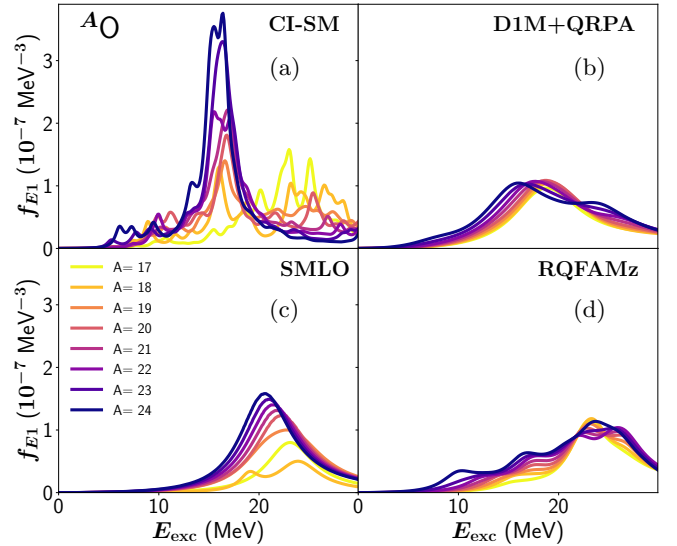
\begin{figure}[htbp]
    \begin{center}
       \centering
\includegraphics[width=0.48\textwidth]{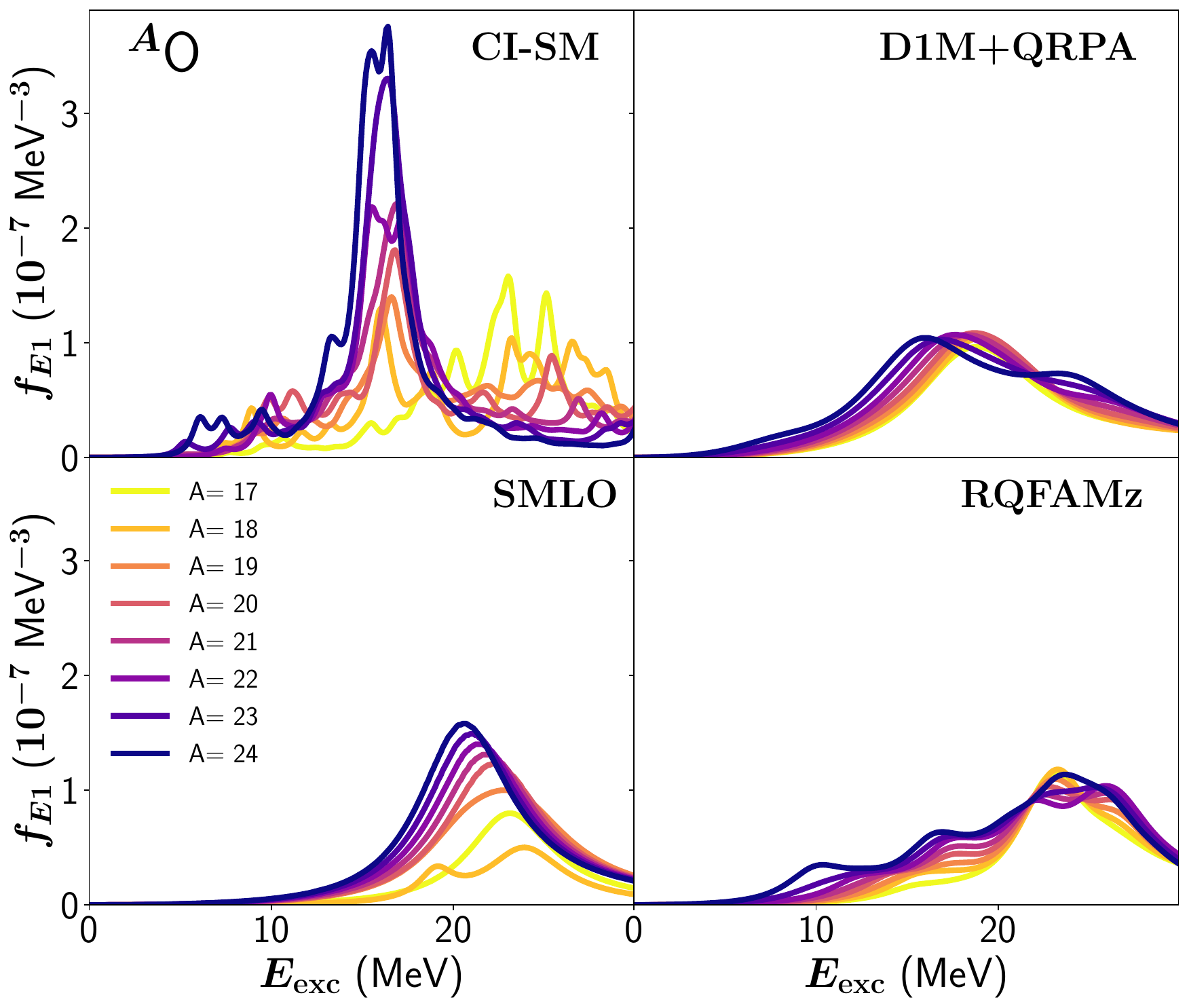}
    \begin{tikzpicture}[overlay]
        \node[anchor=north west] at (-1.0,7.0) {(a)};
        \node[anchor=north west] at (3.0,7.0) {(b)};
        \node[anchor=north west] at (-1.0,3.8) {(c)};
        \node[anchor=north west] at (3.0,3.8) {(d)};
    \end{tikzpicture}
\caption{PSFs for the oxygen isotopes with $N$ ranging between 9 and 16. Panel (a): CI-SM. Panel (b) D1M + QRPA. Panel (c): SMLO. Panel (d): RQFAMz. }
\label{OAgrid}
\end{center}
\end{figure}

\begin{figure}[htbp]
    \begin{center}
       \centering
\includegraphics[width=0.48\textwidth]{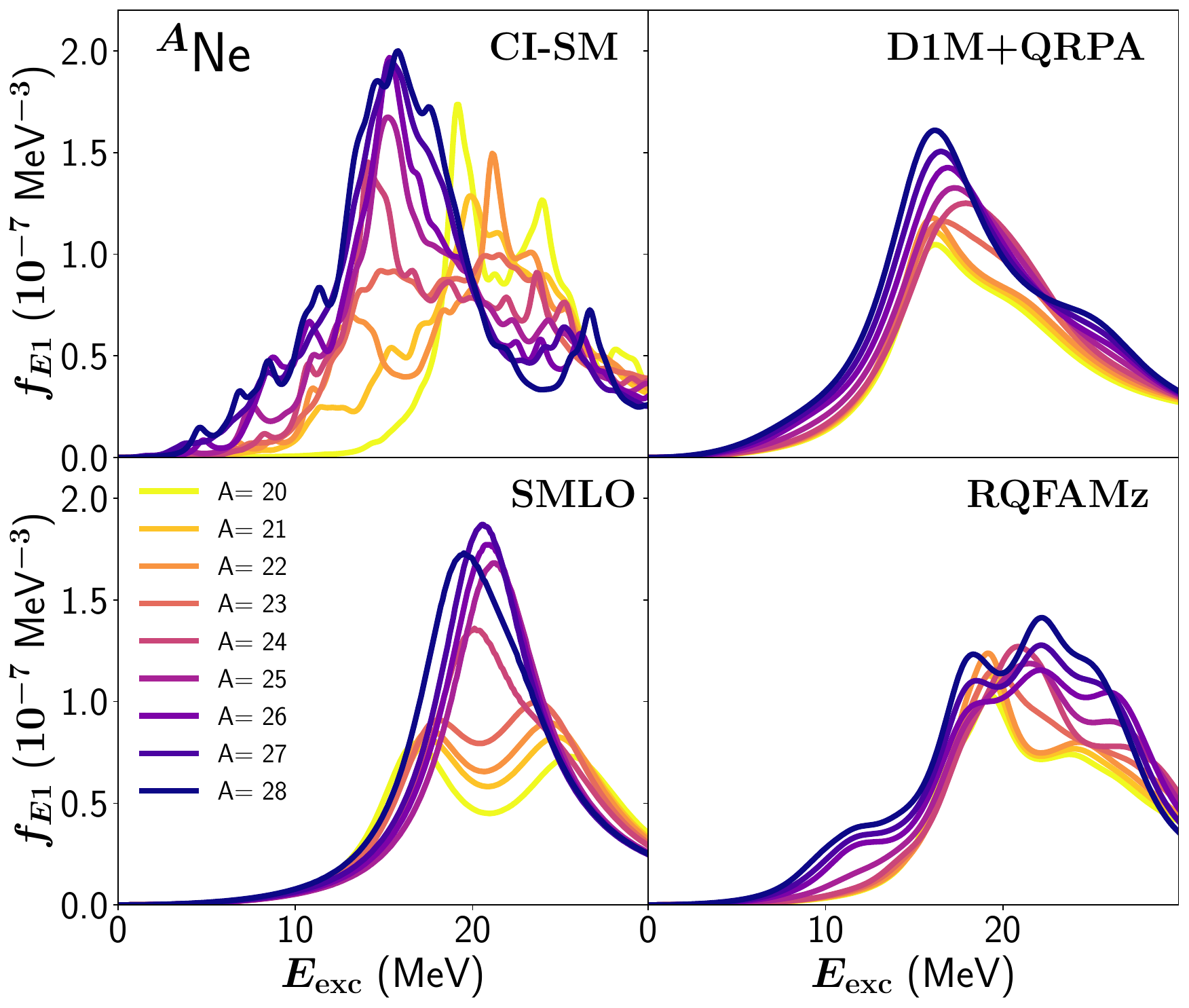}
    \begin{tikzpicture}[overlay]
        \node[anchor=north west] at (-0.5,7.0) {(a)};
        \node[anchor=north west] at (3.0,7.0) {(b)};
        \node[anchor=north west] at (-0.5,3.8) {(c)};
        \node[anchor=north west] at (3.0,3.8) {(d)};
    \end{tikzpicture}
\caption{PSFs for the neon isotopes with $N$ ranging between 9 and 18. Panel (a): CI-SM. Panel (b) D1M + QRPA. Panel (c): SMLO. Panel (d): RQFAMz.}
\label{NeAgrid}
\end{center}
\end{figure}

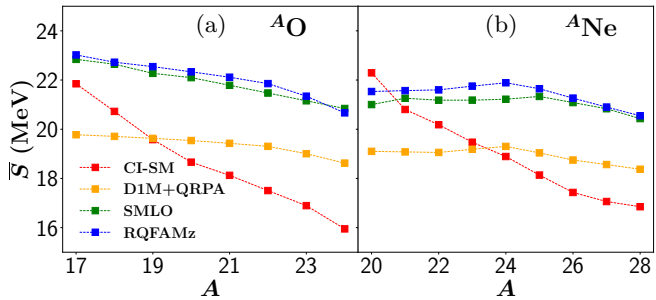
\begin{figure}[htbp]
    \begin{center}
       \centering
    % \begin{subfigure}[b]{0.23\textwidth}
    %     \includegraphics[width=\textwidth]{Ne/Centro_A_.pdf}
    %                             \begin{tikzpicture}[overlay]
    %         \node[anchor=north west] at (1.1,3.7) {(a)};
    %     \end{tikzpicture}
    % \end{subfigure}
    % \hfill
    %     \begin{subfigure}[b]{0.23\textwidth}
    %     \includegraphics[width=\textwidth]{O/Centro_A_.pdf}
    %                                     \begin{tikzpicture}[overlay]
    %         \node[anchor=north west] at (1.1,3.7) {(b)};
    %     \end{tikzpicture}
    % \end{subfigure}
    \includegraphics[width=0.48\textwidth]{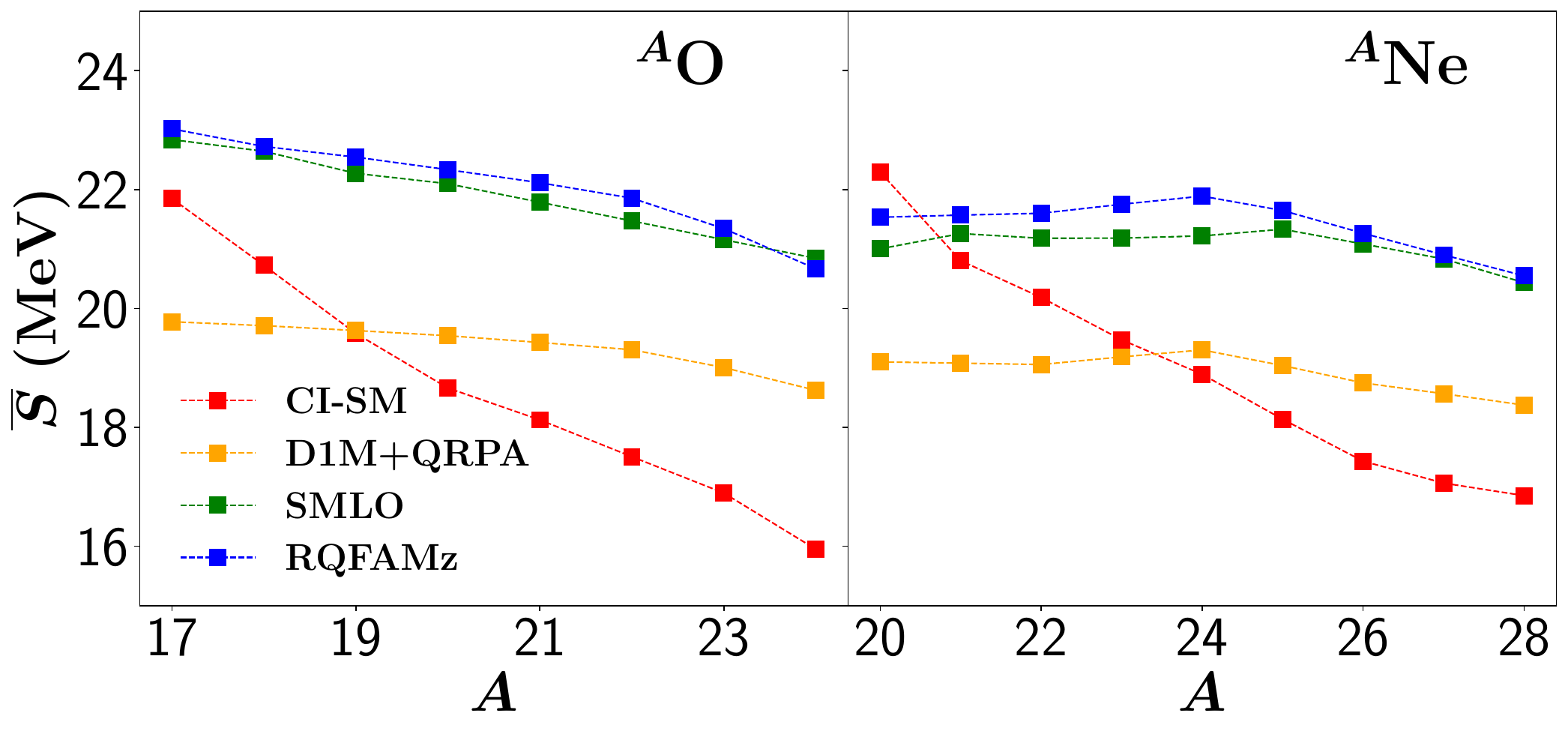}
                \begin{tikzpicture}[overlay]
        \node[anchor=north west] at (-1.85,4.35) {(a)};
    \end{tikzpicture}
                    \begin{tikzpicture}[overlay]
        \node[anchor=north west] at (1.85,4.35) {(b)};
    \end{tikzpicture}
    \caption{Centroids (up to $30$ MeV) of $E1$ distributions as a function of the mass number $A$ for the oxygen isotopic chain (panel (a)) and the neon isotopic chain (panel (b)). In red our CI-SM predictions, in orange D1M+QRPA \cite{goriely_reference_2019}, in blue RQFAMz \cite{gonzalez-miret_zaragoza_large-scale_2025} and in green SMLO results \cite{plujko_giant_2018, goriely_simple_2019}.}    
\label{centro_A}
\end{center}
\end{figure}

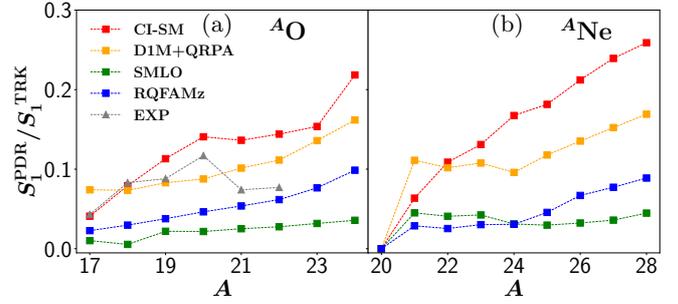
\begin{figure}[htbp]
    \begin{center}
       \centering
    % \begin{subfigure}[b]{0.23\textwidth}
    %     \includegraphics[width=\textwidth]{Ne/S1_pdr_A_.pdf}
    %                                     \begin{tikzpicture}[overlay]
    %         \node[anchor=north west] at (1.1,3.4) {(a)};
    %     \end{tikzpicture}
    % \end{subfigure}
    %         \hfill
    % \begin{subfigure}[b]{0.23\textwidth}
    %     \includegraphics[width=\textwidth]{O/S1_pdr_A_.pdf}
    %                                     \begin{tikzpicture}[overlay]
    %         \node[anchor=north west] at (1.1,3.4) {(b)};
    %     \end{tikzpicture}
    % \end{subfigure}
        \includegraphics[width=0.48\textwidth]{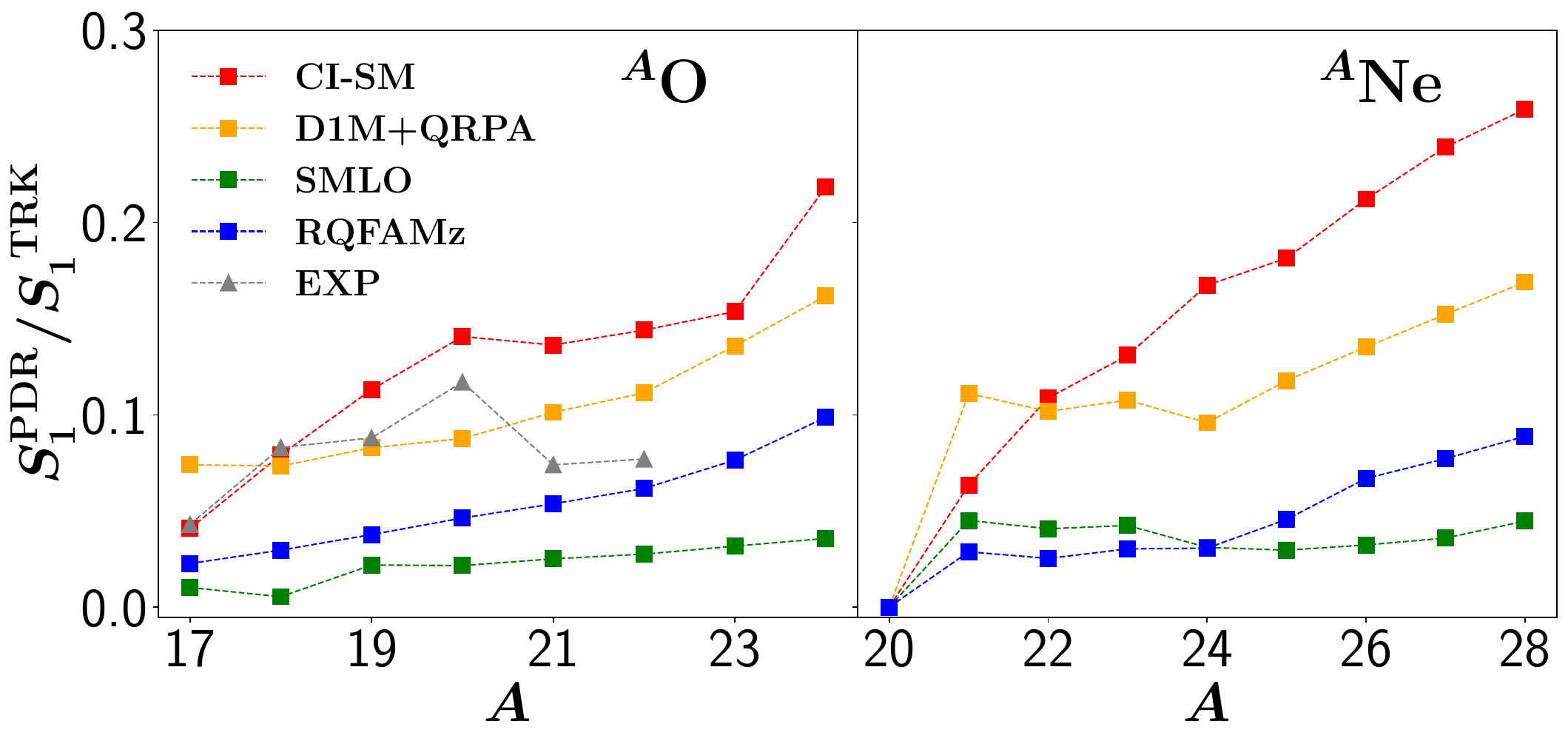}
                        \begin{tikzpicture}[overlay]
        \node[anchor=north west] at (-1.85,4.35) {(a)};
    \end{tikzpicture}
                    \begin{tikzpicture}[overlay]
        \node[anchor=north west] at (1.85,4.35) {(b)};
    \end{tikzpicture}
    \caption{Panel (a): fraction of the energy weighted sum rule in the PDR region (from the neutron separation energy $S_n$ to $15$ MeV) of $E1$ distributions as a function of the mass number $A$ for the oxygen isotopic chain. Panel (b): same for the neon isotopic chain. In red our CI-SM predictions, in orange D1M+QRPA \cite{goriely_reference_2019}, in blue RQFAMz \cite{gonzalez-miret_zaragoza_large-scale_2025}, in green SMLO results \cite{plujko_giant_2018, goriely_simple_2019} and in grey experimental values from Ref. \cite{leistenschneider_photoneutron_2001}.}    
\label{pdr_A}
\end{center}
\end{figure}

After comparing nuclei close to the stability line, we turn to the predictions in the neutron-rich region, where nuclear models tend to show the largest discrepancies. In Figs.~\ref{OAgrid} and \ref{NeAgrid}, we compare the PSF predictions from all four models for oxygen and neon isotopic chains. As observed earlier, the CI-SM PSFs exhibit significantly more fragmentation than the linear-response-based approaches. While D1M+QRPA and RQFAMz show a smooth evolution of the PSFs with increasing neutron number, the CI-SM dipole responses display sharper variations. Clearly, the SMLO predictions were not tailored for neutron-rich nuclei and therefore cannot account for any PDR. 
% Consequently, the rightmost lower panels of Figs.~\ref{Ne_A_psf} and \ref{O_A_psf} represent only the GDR tails. 
In all three microscopic models, we observe an accumulation of low-lying dipole strength with increasing neutron number, characteristic of the emergence of a PDR. Although it may seem at first sight that no PDR appears in the D1M+QRPA case, it is, in fact, dissolved in the GDR tail. 
% , as becomes apparent on a logarithmic scale (see panels (e) to (h) in Figs.~\ref{Ne_A_psf} and \ref{O_A_psf}). 
In Refs.~\cite{sieja23,le_noan_configuration_2025}, we showed that the CI-SM prediction for the PDR in $^{26}$Ne is very close to the experimental values (both strength and centroid), which is promising for extending such predictions to other neutron-rich nuclei. Fig.~\ref{centro_A} shows the centroids (calculated up to $30$ MeV) for all four models. As can be seen, D1M+QRPA, RQFAMz, and SMLO exhibit the same trend as a function of neutron excess, displaying a gradual decrease. The difference in absolute values between D1M+QRPA and RQFAMz can be attributed to the fact that RQFAMz predicts significantly more strength at high energies ($20$–$30$ MeV) than D1M+QRPA. Additionally, phenomenological corrections made in D1M+QRPA PSFs optimize the centroid energies in medium and heavy mass nuclei but tend to underestimate them for light nuclei \cite{goriely_gogny-hfbqrpa_2018}. 
Interestingly, the CI-SM centroids follow a very different trend from the other models, exhibiting a much steeper decrease with increasing $A$. This effect is also evident in Fig.~\ref{pdr_A}, where we show the fraction of the PDR EWSR calculated up to $15$ MeV relative to the TRK sum rule\footnote{The choice of $15$~MeV as the upper bound, rather than $10$~MeV, is motivated by the fact that the experimental values in Ref.~\cite{leistenschneider_photoneutron_2001} are reported up to this energy. As a consequence, this strength is not purely PDR since it incorporates part of the GDR tail.}. Clearly, our calculations predict a stronger increase in the low-lying strength as a function of neutron excess. For the oxygen isotopic chain, experimental data on the PDR are available for a few isotopes~\cite{leistenschneider_photoneutron_2001}. As can be seen in the left panel of Fig.~\ref{pdr_A}, the CI-SM predictions agree quite well with the experiment up to $^{20}$O. While they deviate for $^{21}$O and $^{22}$O, the overall trend remains consistent with the data. This deviation can be attributed to the systematic underestimation of the PSF centroid discussed in the previous section. Indeed, introducing a $1$ MeV upward shift would lead to a much better agreement with the data. While the differences in nuclear models observed for neutron-rich nuclei appear substantial, they have no impact on the modeling of the UHECR, which is dominated by the description of stable and proton-rich nuclei (see Sec.~\ref{sec:uhecr}). Similar model-to-model variations also appear on the proton-rich side, although the proximity of the proton drip line suppresses the most divergent cases.

\section{Application to UHECR propagation\label{UHECR}}
\label{sec:uhecr}

Recent analyses of the data from UHECR yielded important information, such as the contribution of heavy nuclei in their composition and the anisotropy of the source
\cite{aab17,all22,all24}. The importance of the $E1$ PSF in such process has been recently stressed by the advent of the PANDORA collaboration, aiming at a better determination of the $E1$ PSF both from experiment and theory \cite{tamii_pandora_2022}. Here, we illustrate the impact on the UHECR propagation of various theoretical approaches for the $E1$ PSF, including phenomenological ones, such as SMLO, and microscopic D1M+QRPA, RQFAMz, and CI-SM ones. It should be noted that the infra-red background is also involved in the propagation of the UHECR \cite{all22,tamii_pandora_2022}, but focusing on the cosmic microwave background (CMB) already allows evaluating the impact of various $E1$ PSF models. 

The propagation of nuclei in the intergalactic medium is
influenced by interactions with the background radiation fields 
through photodisintegration reactions. In the nucleus rest frame, at typical UHECR energies of 10$^{19}$-10$^{21}$~eV, the CMB photons are boosted to energies in the range between a few
hundreds of keV up to a few hundreds of MeV. The interaction process between
the UHECRs and the CMB is dominated by the GDR at
photon energies below 30-50~MeV, and to a lesser extent by the quasideuteron
emission for intermediate energies (between 50~MeV and 150~MeV) and the pion
photoproductions at energies above 150~MeV (see Ref.~\cite{kha05} and references therein). 

To describe the changes in abundance of the heavy nuclei as
a result of the interaction of the UHECR with the CMB, a nuclear reaction network including
all interactions of interest must be considered. The chosen set
of nuclear species are coupled by a system of differential equations
corresponding to all the reactions affecting each nucleus, i.e. mainly
photodisintegrations and $\beta$-decays. The rate of change of the number density
$N_{Z,A}$ of a nucleus $(Z,A)$ with charge number $Z$ and mass number $A$ can   
be written as

\begin{eqnarray}
\label{eq1}
{dN_{Z,A}  \over dt} &  & =  N_{Z+1,A} \lambda_{\beta}^{Z+1,A} + N_{Z-1,A} \lambda_{\beta}^{Z-1,A} \nonumber \\ 
& &+ N_{Z,A+1} \lambda_{\gamma,n}^{Z,A+1} + N_{Z+1,A+1} \lambda_{\gamma,p}^{Z+1,A+1} \nonumber \\
& & + N_{Z+2,A+4} \lambda_{\gamma,\alpha}^{Z+2,A+4} + N_{Z,A+2} \lambda_{\gamma,2n}^{Z,A+2} \nonumber \\
& &  + N_{Z+2,A+2} \lambda_{\gamma,2p}^{Z+2,A+2}  + N_{Z+4,A+8} \lambda_{\gamma,2\alpha}^{Z+4,A+8}  \nonumber \\
& &+ N_{Z+1,A+2} \lambda_{\gamma,np}^{Z+1,A+2}  + N_{Z+2,A+5}\lambda_{\gamma,n\alpha}^{Z+2,A+5}  \nonumber \\
& &+  N_{Z+3,A+5} \lambda_{\gamma,p\alpha}^{Z+3,A+5} \nonumber  \\  
& & - N_{Z,A} \left\lbrack\lambda_{\beta}^{Z,A}+ \sum_x \lambda_{\gamma,x}^{Z,A} \right\rbrack~, 
\end{eqnarray}

\noindent where $\lambda_{\beta}^{Z,A}$ is the $\beta$-decay rate of nucleus $(Z,A)$ and
$\lambda_{\gamma,x}^{Z,A}$ its photodisintegration rate followed by the
emission of a single neutron ($x=n$), proton ($x=p$) or  $\alpha$-particle ($x=\alpha$) or
the emission of multiple particles such as $2n$, $2p$, $2\alpha$, $np$, $\dots$,
including all open channels for a given photon energy distribution.

The CMB photon density n($\epsilon$) depends only on the UHECR Lorentz
factor $\gamma=E/Mc^{2}$ (where $E$ is the UHECR energy, $M$ its mass, and $c$ the light speed)
\cite{st99}. The calculations of the CMB density as a function of the photon
energy $\epsilon$ in the nucleus rest frame show that photon energies
overlap with the nuclear GDR for $\gamma$ ranging from $5 \times 10^9$ to
10$^{12}$. In the nucleus rest frame, the photodisintegration rate
$\lambda_{\gamma,x}$ can be estimated from the cross section
$\sigma_{\gamma,x}(\epsilon)$ by
\begin{equation}
\lambda_{\gamma,x}=\int{n(\epsilon)~\sigma_{\gamma,x}(\epsilon)~c~d\epsilon}~.
\label{eq2}
\end{equation}
The photonuclear cross sections $\sigma_{\gamma,x}(\epsilon)$ are calculated with the TALYS reaction code \cite{kon23} making use of the $E1$ PSF discussed in previous sections. In particular, in the Hauser-Feshbach approach \cite{haus}, $\sigma_{\gamma,x}(\epsilon)\sim T_\gamma. T_x /T_{\rm tot}$ can be expressed in terms of the transmission coefficients $T$, where the photo-transmission coefficient $T_\gamma$ is calculated from the convolution of the PSF and the level density. The level density and the additional input quantities such as the optical potential required to calculate the particle $T_x$  and total $T_{\rm tot}$ transmission coefficients, are taken from the default models in the TALYS code \cite{kon23}.

All nuclei lighter than the seed nuclei and located between the valley of
$\beta$-stability and the proton drip line must be included in the network. In the present case, 
we choose to consider the propagation of a seed $^{40}$Ca nucleus: although large-scale calculation over parts of the nuclear chart is a numerically challenging task for CI-SM approaches, the present CI-SM allows to calculate the $E1$ strengths of all necessary nuclei until mass 40. It should be noted that the interaction of UHECRs with the CMB is expected to include all possible nuclei resulting from the photodisintegration of the
heaviest species and therefore involve all stable and neutron-deficient unstable isotopes with $A\la 56$ \cite{all22}.

\begin{figure}[h]
\centering
% \vspace*{-7mmcm}       % Give the correct figure height in cm
\scalebox{1.1}{\includegraphics[width=8cm,clip]{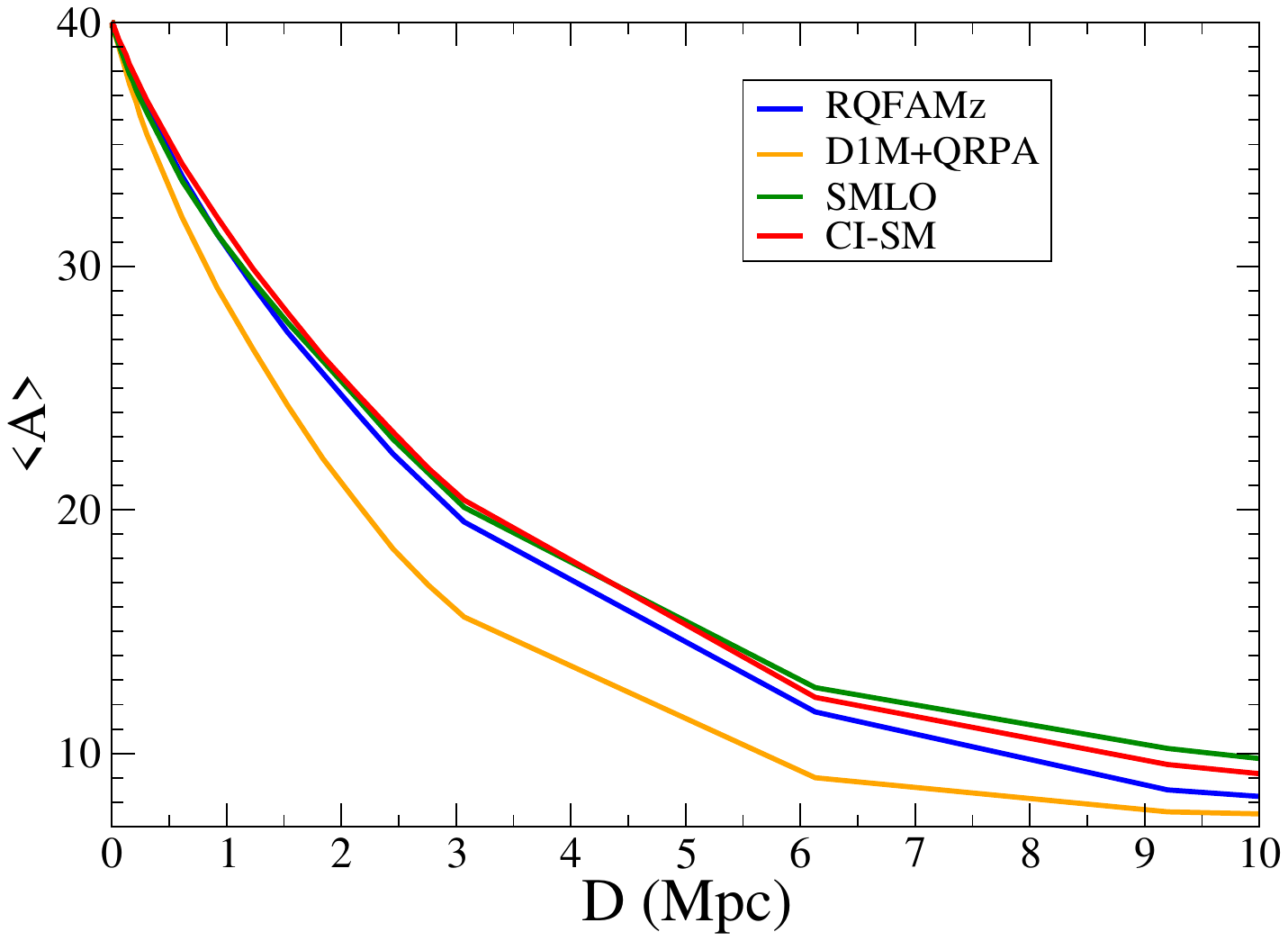}}
\caption{Average value of mass number for the propagation of a source of $^{40}$Ca of 3.7 10$^{20}$ eV as a function of the distance $D$ from the source.}
\label{amoy}       % Give a unique label
\end{figure}

Figure~\ref{amoy} shows the evolution of the mean value of the mass number during the propagation of a source of $^{40}$Ca obtained with cross sections based on CI-SM PSFs. It is compared
to the result based on the D1M+QRPA cross sections \cite{goriely_gogny-hfbqrpa_2018}, the RQFAMz ones \cite{gonzalez-miret_zaragoza_large-scale_2025}, and the phenomenological SMLO approach \cite{plujko_giant_2018}. The CI-SM calculations predict a propagation behavior close to the SMLO and the RQFAMz ones. The D1M+QRPA calculation leads to a significantly smaller survival distance for nuclei than the ones from other models. A shorter propagation distance in the case of the D1M+QRPA calculations implies that the overlap (Eq.~\ref{eq2}) between the CMB photon density and the $E1$ PSF 
is larger than in the CI-SM, SMLO and RQFAMz cases. This is due to the lower energy centroid of the $E1$ strength function in nuclei close to the stability
predicted in the D1M+QRPA case, compared to other calculations, as shown Fig. \ref{SDSsd}a.

\begin{figure}[h]
\centering
% \vspace*{-1cm}       % Give the correct figure height in cm
\scalebox{1.1}{\includegraphics[width=8cm,clip]{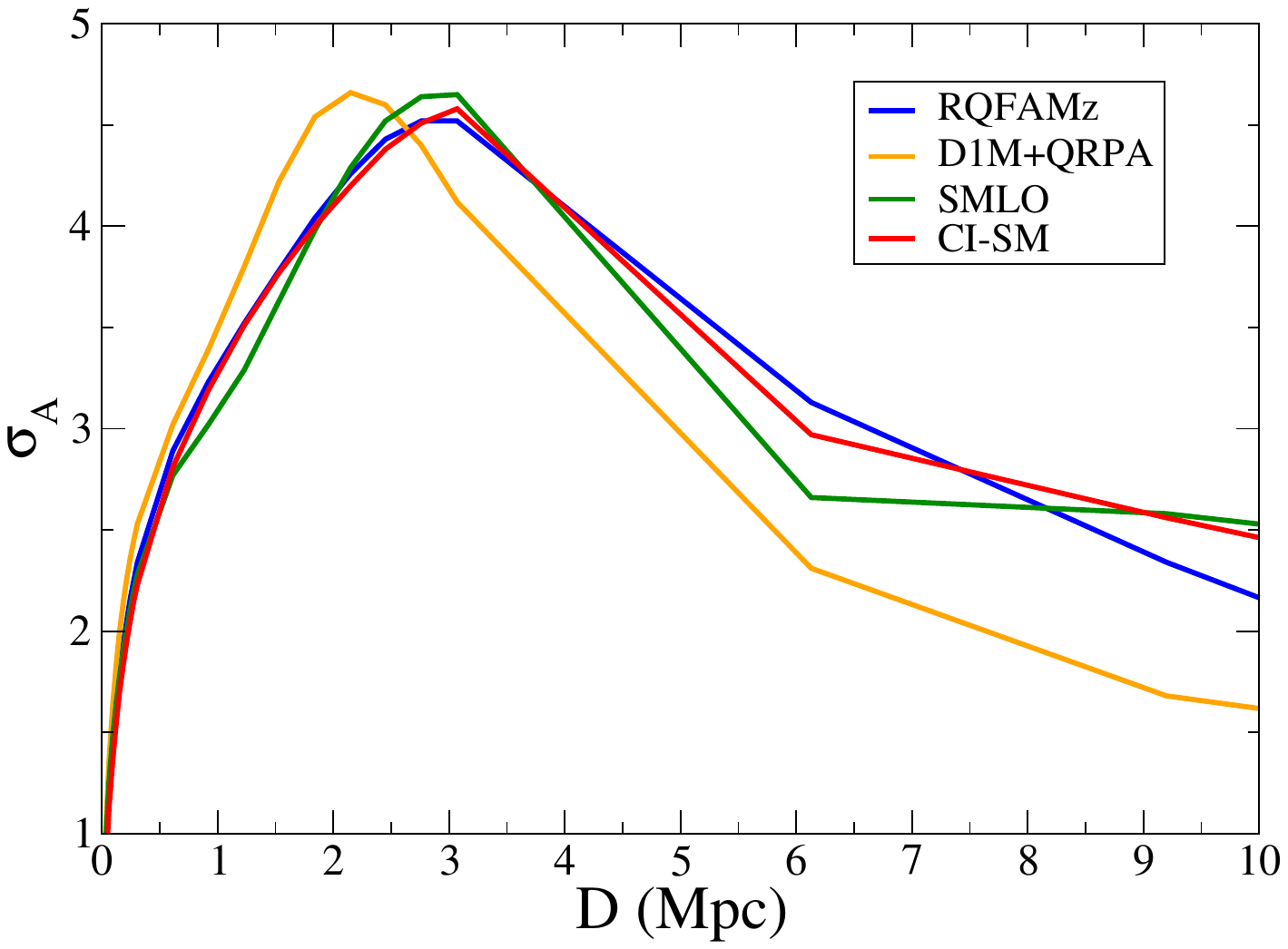}}
\caption{Dispersion around the mean mass value as a function of the distance $D$ from the source.}
\label{disp}       % Give a unique label
\end{figure}

Similar conclusions are obtained on the dispersion of the mass number during the propagation of the UHECR, shown in Fig.~\ref{disp}, where the D1M+QRPA prediction deviates most from the remaining models. The dispersion increases first in the case of the D1M+QRPA calculation but falls below the other models after 3~Mpc. The CI-SM follows closely RQFAMz and SMLO at first but predicts different dispersions beyond 5~Mpc.
%One can conclude from the present results that the interaction of the UHECR with CMB is largely dominated by 
%the theoretical description of the position of the centroid close to the stability linex. 

\section{Conclusion\label{CONC}}
We have performed a systematic evaluation of $E1$ photoresponse of nuclei with masses $A=6-40$ within the CI-SM framework building on earlier studies from Ref.~\cite{le_noan_electric_2025}. The results were compared to experimental data when available and to other recommended sets of PSF, namely the traditional SMLO approach and predictions based on two microscopic approaches, D1M+QRPA and RQFAMz. The CI-SM results, as expected, exhibit more fragmentation 
than other models and less smooth behavior with mass. In neutron-rich nuclei, the CI-SM predictions lead to more $E1$ strength accumulated
at low energies and thus to lower centroids. The agreement of CI-SM with available experimental data is slightly better than for other models, in spite of any phenomenological adjustments on the computed $E1$ strength centroid or width.  
While the CI-SM predicts stronger structure effects, such as a splitting of the GDR in some nuclei, in contrast to other models, generally the centroids of the CI-SM PSF are closer to SMLO and RQFAMz predictions than to the D1M+QRPA ones. This is further reflected in the UHECR
propagation distance of a $^{40}$Ca source calculated with the use of PSF from those models, where the SMLO, RQFAMz and CI-SM results of average mass value of UHECR are relatively close, but clearly differ from the D1M+QRPA predictions.

It should be noted that nuclei up to $^{56}$Fe are expected to be present in the UHECR source; thus the CI-SM predictions of this work covered only $\sim 50\%$ of nuclei necessary for the complete modeling of UHECR. Based on the previous CI-SM calculations of \cite{Sieja-PRL}, we are currently developing a suitable approach to also address dipole response of $pf$-shell nuclei. The current diagonalization limit of $\sim 10^{10}$ will allow us to compute the $E1$ strength function of $pf$-shell nuclei up to $^{50}$Fe, without any further approximation, and thus to fill in further the existing gap of CI-SM predictions of relevance. While here we examined how the PSF model may affect the UHECR interaction with the CMB, it was demonstrated \cite{Boncioli2017} that the impact of nuclear model uncertainties is potentially larger in environments with non-thermal radiation fields, e.g. in the possible astrophysical sources of UHECR such as the gamma-ray bursts. Such sensitivity studies could be performed in the future with TALYS cross sections based on the microscopic CI-SM predictions.

\section{Acknowledgments}
OLN and KS acknowledge support of the Interdisciplinary Thematic Institute QMat, as part of the ITI 2021-2028 program of the University of Strasbourg, CNRS and Inserm, supported by IdEx Unistra (ANR 10 IDEX 0002), and by SFRI STRAT’US project (ANR 20 SFRI 0012) and EUR QMAT ANR-17-EURE-0024 under the framework of the French Investments for the Future Program. SG acknowledges financial support from F.R.S.-FNRS (Belgium). This work was supported by the Fonds de la Recherche Scientifique - FNRS and the Fonds Wetenschappelijk Onderzoek - Vlaanderen (Belgium) under the EOS Project No O000422.

\bibliographystyle{apsrev}
\bibliography{main}

\end{document}